\documentclass[fleqn,usenatbib]{aa}

\usepackage{placeins}
\usepackage{graphicx}
\usepackage{amsmath, amssymb}
\usepackage{txfonts}
\usepackage{natbib}
\usepackage{xspace}
\usepackage{upgreek}
\usepackage{xcolor}
\bibpunct{(}{)}{;}{a}{}{,}
\usepackage{hyperref}
\hypersetup{
    colorlinks = true,
    citecolor  = blue,  
    linkcolor  = blue,  
    urlcolor   = blue    
}

\newcommand{\pc}{\,\mathrm{pc}}
\newcommand{\Msun}{\,\mathrm{M}_{\odot}}
\newcommand{\kpc}{\,\mathrm{kpc}}
\newcommand{\Gyr}{\,\mathrm{Gyr}}

\newcommand{\kms}{\,\mathrm{km\,s}^{-1}}

\newcommand{\Rg}{R_\mathrm{g}}
\newcommand{\Nbody}{$N$-body\xspace}

\newcommand{\Adep}{\mathrm{A}_\mathrm{DEP}}
\newcommand{\Aout}{\mathrm{A}_\mathrm{OUT}}
\newcommand{\Pdep}{\mathrm{P}_\mathrm{DEP}}

\newcommand{\Pfdep}{\mathrm{P}_\mathrm{flip,DEP}}

\newcommand{\PDFf}{\mathrm{PDF}_\mathrm{flip}}
\newcommand{\PRdep}{\overset{\large \sim}{{\mathrm{P}}}_\mathrm{DEP}}
\newcommand{\PRout}{\overset{\large \sim}{{\mathrm{P}}}_\mathrm{OUT}}
\newcommand{\NNDRdep}{\overset{\large \sim}{{\mathrm{NND}}}_\mathrm{DEP}}
\newcommand{\NNDRout}{\overset{\large \sim}{{\mathrm{NND}}}_\mathrm{OUT}}

\newcommand{\pksdep}{p_\mathrm{KS-DEP}}
\newcommand{\pksout}{p_\mathrm{KS-OUT}}
\newcommand{\pchi}{p_\mathrm{\upchi^2}}
\newcommand{\ppoisson}{p_\mathrm{Poisson}}

\newcommand{\secref}[1]{Section~\ref{#1}}
\newcommand{\figref}[1]{Figure~\ref{#1}}
\newcommand{\tabref}[1]{Table~\ref{#1}}

\graphicspath{{Figures/}}

\defcitealias{Rostami2024-metallicity}{Paper~I}

\begin{document}

   \titlerunning{Spatial distribution of GCs in host galaxies}
   \authorrunning{Rostami-Shirazi et al.}

   \title{Universal depletion of metal-poor globular clusters in inner galaxy regions: Fossil record of black hole retention}

   \author{
        Ali Rostami-Shirazi\inst{1}\thanks{E-mail: ali.rostami.shirazi@ipm.ir} \and
        Narges Rostami\inst{2} \and
        Hosein Haghi\inst{1,3,4} \and
        Akram Hasani Zonoozi\inst{3,4}
    }

    \institute{
     School of Astronomy, Institute for Research in Fundamental Sciences (IPM), PO Box 19395-5531, Tehran, Iran
         \and
             Department of Physics, Faculty of Physics, Alzahra University, Vanak, 1993891176, Tehran, Iran
         \and
             Department of Physics, Institute for Advanced Studies in Basic Sciences (IASBS), 444 Prof. Sobouti Blvd., Zanjan 45137-66731, Iran
         \and
             Helmholtz-Institut f\"ur Strahlen-und Kernphysik (HISKP), Universit\"at Bonn, Nussallee 14-16, D-53115 Bonn, Germany
             }

   \date{Received XXX; accepted YYY}

  \abstract{We analyzed the spatial distribution of globular cluster (GC) systems across 37 host galaxies in a two-dimensional parameter space defined by projected galactocentric distances ($\Rg$) and metallicity ([Fe/H]). We identified a universal triangular depleted region, characterized by a lack of metal-poor GCs in the inner parts of host galaxies. The morphology of this depleted region correlates with the luminous mass of host galaxies; more massive galaxies consistently exhibit more extended depleted regions. We attribute this phenomenon to the combined influence of large-scale galactic assembly and internal GC dynamics, particularly the initial retention of black holes (BHs) within GCs. Metal-poor GCs harbor a more massive and compact BH subsystem, which fosters more energetic few-body encounters, injecting greater kinetic energy into the stellar population. This extra energy, combined with the strong tidal forces in the galactic central regions, accelerates the dissolution of lower-metallicity GCs on timescales shorter than the host galaxy’s age, leading to the emergence of a triangular depleted pattern in the $\Rg$-[Fe/H] parameter space. Stronger tidal fields in more massive galaxies confine surviving metal-poor GCs to larger radii, broadening the depleted region. The depleted region's morphology may serve as a potential distance indicator for host galaxies. Our results also suggest that scenarios involving substantial BH natal kicks are unlikely, as too few retained BHs would erase the metallicity-dependent cluster dissolution required to form the observed depletion region.}


   \maketitle

\section{Introduction}\label{sec:intro}

The bimodal metallicity distribution of globular clusters (GCs) in most large host galaxies, such as the Milky Way (MW), implies the existence of two distinct subpopulations: metal-poor and metal-rich GCs \citep{Zinn1985,Forbes1997,Peng2006,Chies-Santos2012,Harris2023,Hartman2023}. These subpopulations exhibit dissimilarities in spatial distribution \citep{Bassino2006,Peng2008,Strader2011,Pota2013}, kinematics \citep{Schuberth2010,Strader2011,Pota2013}, age \citep{Hansen2013}, and size \citep{Kundu1998,jordan2004,Harris2009,Strader2012} within their host galaxies, suggesting divergent formation mechanisms. The metal-rich GC subpopulation is believed to have formed concurrently with the majority of the galaxy’s stellar component, whereas the metal-poor GCs are thought to be captured by the host galaxy through interactions with satellite galaxies \citep{Forbes1997,cote2000,LAW2010b,Tonini2013,Pfeffer2018,massari2019,Kruijssen2020,Rostami2022,Shirazi2023}.

Observational data from the MW GC system indicate that populations with different metallicities occupy distinct radial distributions. In particular, the inner part of the MW is almost depleted of metal-poor GCs \citep{Harris2001,Bica2006}. The lack of metal-poor GCs at low mean Galactocentric distances (semi-major axes; $a$) follows a triangular pattern in the $a$-[Fe/H] diagram (see Figure 1 in \citealt{Rostami2024-metallicity}, hereafter \citetalias{Rostami2024-metallicity}), indicating that GCs with lower metallicities appear further away from the Galactic center. Density analysis at a larger $a$, where both metal-poor and metal-rich GCs coexist, reveals that metal-rich GCs are significantly denser than their metal-poor counterparts (\citetalias{Rostami2024-metallicity}). This substantial density discrepancy implies that metal-poor GCs experience weaker gravitational binding, rendering them more vulnerable to tidal stripping and subsequent dissolution, suggesting a plausible explanation for their absence in the Galactic center. In \citetalias{Rostami2024-metallicity}, we interpreted the emergence of the triangular depleted region in the $a$–[Fe/H] diagram over Hubble time as a consequence of the increased vulnerability of metal-poor GCs to tidal destruction near the MW, coupled with the greater resilience of metal-rich GCs against dissolution. This pattern suggests a metallicity-dependent survival rate for GCs in the inner Galactic regions. Direct \Nbody simulations confirm that initial retention of black holes (BHs) in clusters leads to a significant disparity in evaporation rates between metal-poor and metal-rich clusters (\citealt{Banerjee2017,Debatri2022,Gieles2023}; \citetalias{Rostami2024-metallicity}).

Metallicity plays a crucial role in the evolution of massive stars, affecting the mass-loss rate through stellar winds and, consequently, the final mass of their remnants. Metal-poor clusters can host substantially more massive BHs than their metal-rich counterparts, as metal-rich progenitors undergo greater mass loss via stellar winds \citep{Vink2001,vink2005,Belczynski2008,Belczynski2010}. Moreover, the range of masses that convert to BHs is larger in metal-poor clusters \citep{Shanahan2015}. The heavier BH subsystem (BHSub) in metal-poor clusters, leads to stronger few-body encounters that inject increased kinetic energy into the entire stellar population. This results in greater expansion and higher evaporation rates compared to metal-rich clusters (\citealt{Banerjee2017,Debatri2022,Gieles2023}; \citetalias{Rostami2024-metallicity}). As a result, these clusters are more likely to dissolve within Hubble time, especially in the vicinity of the Galactic center, creating an empty triangular region in the $a$-[Fe/H] diagram. While metallicity-dependent BH dynamics can accelerate the disruption of low-metallicity clusters, it is likely only one of several mechanisms shaping the observed radial metallicity distribution of GCs. Large-scale processes such as hierarchical assembly, orbital decay, and tidal stripping in dense galactic environments likely act in concert with internal BH-driven heating.

Upon exhausting all thermonuclear energy sources, massive stars undergo gravitational collapse followed by supernova explosions, resulting in the formation of BHs as remnants \citep{oppenheimer1939continued,iben1984single,heger2003}. The mass and momentum ejected by the supernova are asymmetrically distributed, imparting a natal kick to the compact remnant. Constraining BH natal kicks relies on studying the kinematics of binary systems in the MW where a luminous companion is bound to a BH. However, results remain inconclusive: while some studies report evidence of substantial kicks, others find none \citep{Atri2019,Zhao2023,Dashwood2024,Vigna2024,Nagarajan2025}. The debated magnitude of BH natal kicks creates uncertainties in the initial BH population retained in star clusters. Historically, BHs in GCs were considered unlikely due to the assumption that strong natal kicks could accelerate them beyond the GCs' escape velocity. However, since 2007, observations have confirmed the presence of BHs in GCs, suggesting that their natal kicks are weaker than previously thought \citep{Maccarone2007,Shih2010,Barnard2011,Strader2012BH,Chomiuk2013,Miller-Jones2015,Shishkovsky2018,Giesers2018,Giesers2019,Saracino2022}. Additionally, computational studies have shown that the retention of BHs is crucial to reproduce the observable properties of clusters accurately (\citealt{Merritt2004,Mackey2007,Mackey2008,Morscher2015,Peuten2016,Arca2018,Askar2018,Weatherford2018,Kremer2018,Askar2019,Kremer2019,Weatherford2020,Gieles2021,Torniamenti2023,Dickson2024}; \citetalias{Rostami2024-metallicity}; \citealt{Rostami2024-Remnant,Rostami-Shirazi-UMa}).

This paper focuses on the spatial distribution of GC systems within their host galaxies, specifically investigating whether the triangular region depleted of metal-poor GCs in the inner MW is a universal feature. Detecting this pattern across galaxies could provide additional evidence for the necessity of the initial BH retention within GCs. The paper is organized as follows: \secref{sec:review} outlines the impact of retained BHs on the spatial distribution of GCs, providing context for this study. The results are presented in \secref{sec:results}, followed by a summary and conclusion in \secref{sec:conclusion}.

\section{Impact of retained BHs on GC spatial distribution within host galaxies}\label{sec:review}

\begin{figure}
  \centering
  \includegraphics[width=1\linewidth]{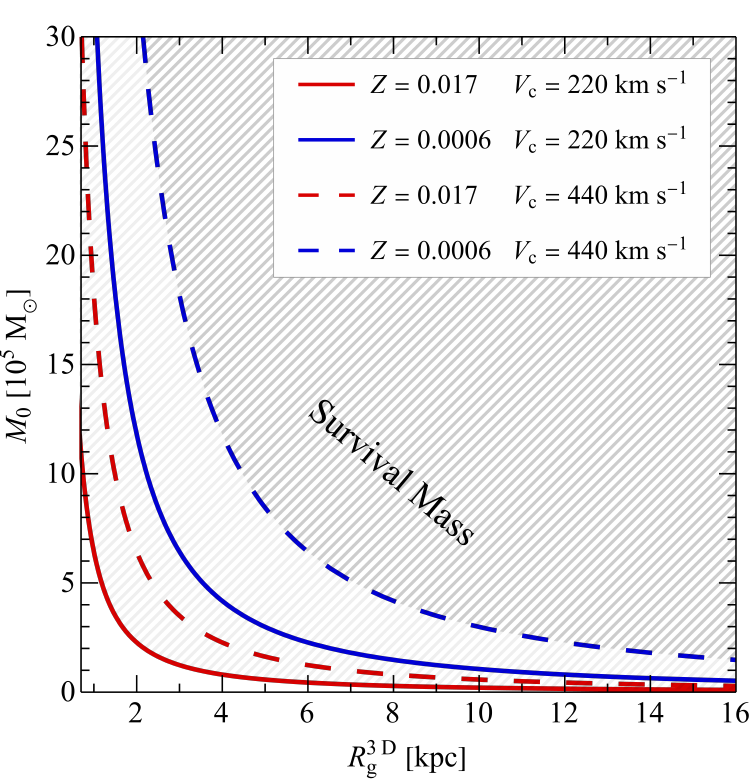}
  \caption{Minimum prestellar evolution mass ($M_0$) required for GCs to survive over Hubble time ($\tau\ge13.8\Gyr$), shown as a function of $R_{\rm g}^{\rm 3D}$. Results are presented for $Z=0.0006$ (blue) and $0.017$ (red) in host galaxies with circular velocities of $V_{\rm c} = 220$ (solid lines) and $440\kms$ (dashed lines). Each curve represents the survival mass threshold; GCs with $M_0$ above the corresponding curve (shaded region) remain bound over Hubble time.}
  \label{fig:SurviveMass}
\end{figure}

\begin{figure}
  \centering
  \includegraphics[width=1\linewidth]{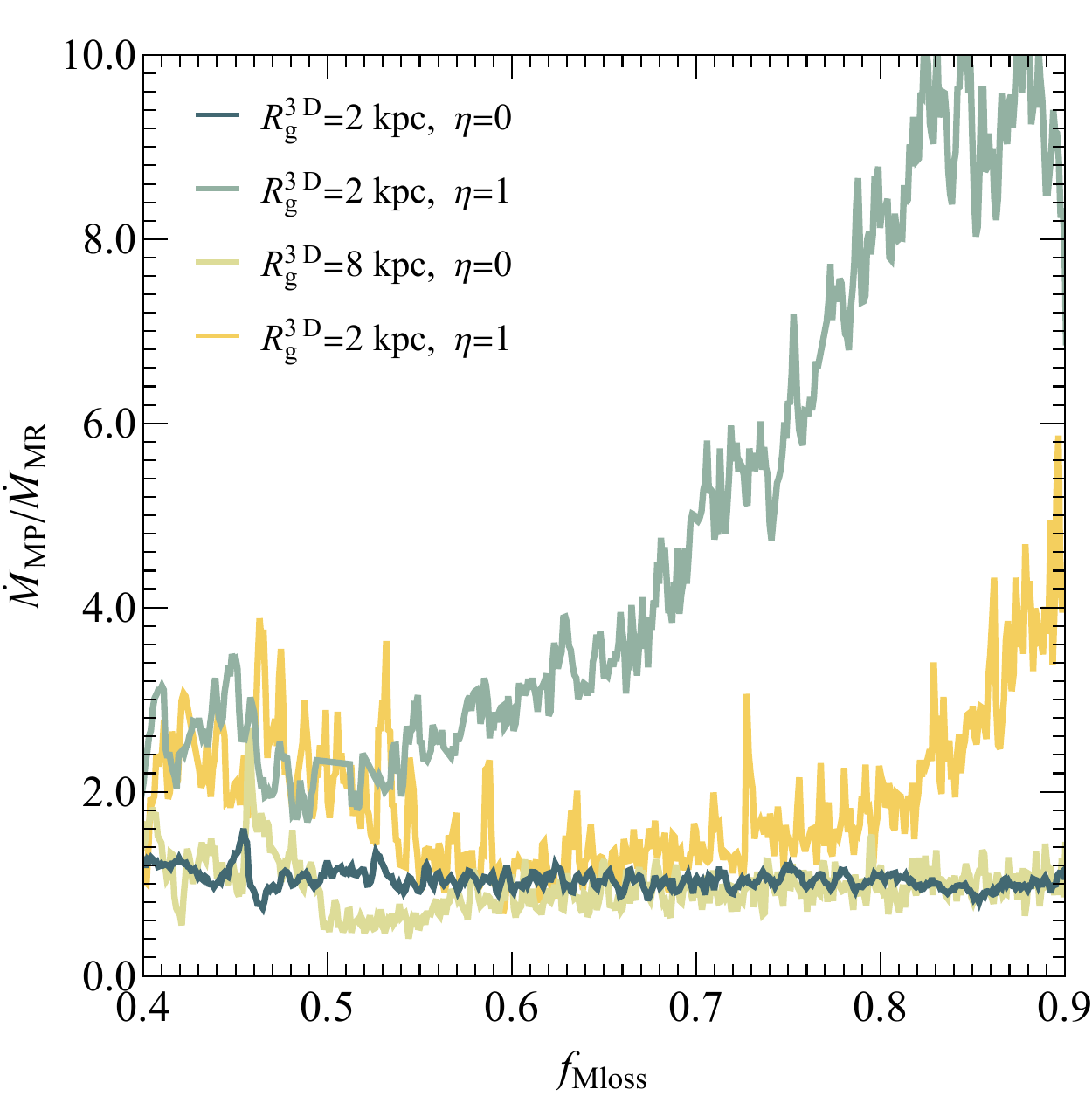}
  \caption{Ratio of metal-poor ($Z=0.0002$) to metal-rich ($Z=0.02$) cluster mass-loss rates, $\dot{M}_{\rm MP}/\dot{M}_{\rm MR}$, as a function of the fractional mass loss, $f_{\rm Mloss}$. The modeled clusters follow the initial conditions of \citetalias{Rostami2024-metallicity}, except for a half-mass radius of $r_{\mathrm{h},0}=1.2 \pc$, and are located at $R_{\rm g}^{\rm 3D}=2$ and $8\kpc$, covering BH retention from complete ejection ($\eta=0$) to full retention ($\eta=1$).}
  \label{fig:Mdot}
\end{figure}

The spatial metallicity distribution of GCs arises from the interplay of large-scale galactic processes and internal GC dynamics. In this context, the initial retention of BHs plays an essential role in driving the depletion of metal-poor GCs in the inner regions of their host galaxies (see \secref{sec:Assembly}). In this section, we employ both literature-derived models (\secref{sec:Literature}) and direct \Nbody simulations (\secref{sec:NBODY}) to examine how BH retention affects the metallicity-dependent dissolution of GCs and consequently the emergence of the depletion pattern.

The retention of a substantial BH population within the cluster after natal kicks triggers the formation of a BHSub at the cluster's center due to Spitzer instability \citep{Spitzer}. This centrally segregated BHSub exhibits intense dynamical activity through subsequent encounters between BH-BH binaries and single BHs, injecting kinetic energy into the surrounding stellar population, which accelerates the cluster's evaporation rate \citep{heggie2003,Breen2013,Banerjee2017,Banerjee2018,Giesers2019,Longwang2020,DSC,Ghasemi2024}. The mass fraction and average mass of BHs increase with decreasing metallicity \citep{Vink2001,vink2005,Belczynski2008, Belczynski2010,Shanahan2015}. On the other hand, metal-rich clusters undergo enhanced stellar mass loss due to stronger stellar winds, which drives core expansion and results in longer core relaxation times \citep{vink2005,Schulman2012,Mapelli2013}. The heavier and more rapidly relaxed BHSub in metal-poor clusters promotes more energetic few-body interactions, injecting more kinetic energy into the stellar population (\citealt{Banerjee2017,Debatri2022,Gieles2023}; \citetalias{Rostami2024-metallicity}).

In the inner galactic environment, the stronger tidal field limits the tidal radius, reducing the escape energy threshold for stars within clusters. This makes metal-poor GCs particularly susceptible to additional energy injection, which can propel stars beyond their escape velocity, dramatically accelerating their evaporation rate and potentially leading to dissolution within timescales shorter than the host galaxy’s age. As galactic radius increases, clusters are subject to weaker tidal forces, resulting in higher escape velocities and a reduced influence of energy generated by the BHSub on the evaporation rates, thereby facilitating the survival of metal-poor GCs. Indeed, increased BHSub-driven heating in lower-metallicity clusters limits their long-term survival to larger three-dimensional galactocentric distances ($R_{\rm g}^{\rm 3D}$). This process gives rise to a nearly void triangular pattern in the $R_{\rm g}^{\rm 3D}$-[Fe/H] parameter space, characterized by a statistical depletion of metal-poor GCs. Stronger tidal forces in more massive galaxies would cause metal-poor GCs to survive only at greater $R_{\rm g}^{\rm 3D}$, thereby extending the triangular depleted region.

\subsection{Literature-derived models}\label{sec:Literature}

\citet{Gieles2023}, using direct \Nbody simulations, extended the mass-loss rate prescription of \citet{Baumgardt2003} by incorporating the dynamical impact of BHs retained within GCs. Their models adopt the rapid supernova mechanism \citep{Fryer2012}, with fallback-dependent natal kicks, such that momentum conservation leads to the retention of $73\%$ ($50\%$) of the BH mass in metal-poor (metal-rich) clusters. They derived a generalized mass-loss rate expressed as
\begin{equation}
\overset{\cdot}{M} = \overset{\cdot}{M}_{\mathrm{ref}} \left( \frac{M}{M_{\mathrm{i}}} \right)^{1 - y} \left( \frac{M_{\mathrm{i}}}{2 \times 10^5\, M_\odot} \right)^{1/3} \left( \frac{\Omega_{\mathrm{tid}}}{0.32\, \mathrm{Myr}^{-1}} \right),
\label{eq:MassLoss}
\end{equation}
where $M_{\mathrm{i}}$ is the post-stellar-evolution initial mass, $\Omega_{\mathrm{tid}}$ quantifies the strength of the external tidal field, $y$ and $\overset{\cdot}{M}_{\mathrm{ref}}$ are free parameters calibrated through \Nbody simulations. Notably, for an initial density within the half-mass radius of $\rho_{\mathrm{h},0} = 300 \Msun~\mathrm{pc}^{-3}$ in metal-rich clusters ($Z=0.017$ and 0.006), the best-fitting parameters are $\overset{\cdot}{M}_{\mathrm{ref}} = -30\Msun~\mathrm{Myr}^{-1}$ and $y = 2/3$, corresponding to a gradual "skiing" mass-loss regime. In contrast, for metal-poor clusters ($Z=0.0006$), the higher BH retention fraction leads to $\overset{\cdot}{M}_{\mathrm{ref}} = -45\Msun~\mathrm{Myr}^{-1}$ and $y = 4/3$, indicative of an accelerated "jumping" disruption mode. Their models assume a fixed initial filling factor, defined as the ratio of the half-mass radius to the effective Jacobi radius, $r_{\mathrm{h},0}/r_{\mathrm{J,eff}} \simeq 0.05$. Although this value is a simplification and real clusters exhibit a range of filling factors and orbital conditions, it serves as a reasonable approximation to the average structural properties of surviving MW GCs.

\figref{fig:SurviveMass} illustrates how the metallicity-dependent mass-loss trends map onto cluster survival by showing the minimum prestellar evolution mass, $M_{\mathrm{0}}$, taken as $2\times M_{\mathrm{i}}$ under the assumption of $\sim 50\%$ stellar evolutionary mass loss. This is the mass required for a GC to survive the Hubble time at a given $R_{\rm g}^{\rm 3D}$, derived via time integration of Equation~\ref{eq:MassLoss}, across different metallicities and host-galaxy tidal field strengths. At each metallicity, the curves delineate the survival mass boundary: clusters above the curve (shaded region) remain long-lived, while those below dissolve within Hubble time. The boundary is strongly metallicity- and tide-dependent, particularly in the inner regions of the host galaxies. Lower–metallicity systems demand larger $R_{\rm g}^{\rm 3D}$ (or higher $M_{\mathrm{0}}$) to survive, with the shift amplified in more massive galaxies (i.e., those with larger circular velocities $V_{\mathrm{c}}$).

As an illustrative case, GCs born with $M_{0}\simeq2\times10^5\Msun$ require $R_{\rm g}^{\rm 3D}\gtrsim6.5\kpc$ to survive in a $V_{\mathrm{c}}=220\kms$ field, whereas their metal-rich counterparts can persist at radii as small as $2\kpc$. In a stronger tidal field ($V_{\mathrm{c}}=440\kms$), the corresponding thresholds shift outward to $\gtrsim13\kpc$ and $\gtrsim4\kpc$, respectively. A complementary perspective is provided at fixed $R_{\rm g}^{\rm 3D}=4\kpc$ in the $V_{\mathrm{c}}=440\kms$ model: clusters with $Z=0.0006$ require $M_0 \gtrsim 1.15 \times 10^6\Msun$ to survive, nearly an order of magnitude higher than the $M_0 \gtrsim 2.1 \times 10^5\Msun$ needed for $Z=0.017$ or $0.006$. At larger $R_{\rm g}^{\rm 3D}$, these offsets diminish as the survival thresholds converge. This pronounced disparity in survival mass at small $R_{\rm g}^{\rm 3D}$ reveals that metal-rich clusters are far less susceptible to disruption in the inner galaxy, whereas metal-poor GCs face significantly stricter survival constraints, leading to a statistical depletion in the $R_{\rm g}^{\rm 3D}$–[Fe/H] parameter space.

\subsection{\Nbody simulations}\label{sec:NBODY}

In \citetalias{Rostami2024-metallicity}, we conducted two sets of \Nbody simulations for both metal-poor and metal-rich models, varying the BH natal kick velocities between low and high. Under the assumption that all BHs are promptly ejected following their formation, we found that the dissolution times of metal-poor and metal-rich clusters were nearly identical, with no evidence for a depleted population of metal-poor clusters in the inner regions of the host galaxy. In contrast, our results indicated that the dissolution time–metallicity correlation is contingent upon the initial retention of BHs within the modeled clusters.

The evolution and eventual self-depletion of the BHSub are governed by the flow of energy through the cluster's half-mass radius, which is primarily regulated by its two-body relaxation time \citep{Breen2013}. In initially compact clusters, short relaxation times drive rapid energy transfer from the BHSub to the stellar component and hence prompt BHSub depletion; by contrast, tidally filling or moderately under-filling systems can expand more effectively, lengthening the relaxation time and prolonging BHSub longevity. Following \citetalias{Rostami2024-metallicity}, we performed an additional suite of \Nbody simulations with a more compact initial configuration, better representing the densities inferred for MW GCs. All models have an initial mass of $M_{\rm 0}=6\times10^4\Msun$. The initial half-mass radius was set to $r_{\mathrm{h},0}=1.2 \pc$, based on the \citet{markskroupa} relation  after gas expulsion, assuming an expansion factor of three \citep{baumgardt2007}. We explore two bracketing metallicities, $Z=0.0002$ and $Z=0.02$, and two limiting prescriptions for BH retention, $\eta \in \{0,1\}$, corresponding to the complete ejection ($\eta=0$) and the full retention of the nascent BH population ($\eta=1$). \figref{fig:Mdot} presents the ratio of mass-loss rates of metal-poor to metal-rich models, $\dot{M}_{\rm MP}/\dot{M}_{\rm MR}$, at $R_{\rm g}^{\rm 3D}=2$ and $8\kpc$, shown as a function of the mass-loss fraction $f_{\rm Mloss}$.

In the $\eta=0$ case, metal-poor and metal-rich clusters lose mass at nearly identical rates throughout their evolution, $\dot{M}_{\rm MP}/\dot{M}_{\rm MR}\approx1$. By contrast, in the $\eta=1$ case, the ratio becomes strongly radius-dependent: at $R_{\rm g}^{\rm 3D}=2\kpc$ it rises steeply, reaching up to $\dot{M}_{\rm MP}/\dot{M}_{\rm MR}=10$, while at $8\kpc$ the excess persists but is more modest. This behavior reflects the interplay between (i) stronger dynamical heating from the heavier BHSub in metal-poor clusters and (ii) the external tide. Near the galaxy center, the reduced Jacobi radius lowers the escape energy, so BH-driven heating is efficiently converted into escapees and an elevated $\dot{M}_{\rm MP}$; farther out, the same heating has a diminished effect because the tidal boundary is wider, and, simultaneously, fractional BH mass gradually decreases due to self-depletion, reducing its contribution to the mass-loss rate. At still larger radii, the ratio asymptotes to unity, and metal-poor and metal-rich clusters evolve similar to BH-free systems after their BHSubs are exhausted.

Crucially, even for these compact initial conditions, metal-poor clusters at small $R_{\rm g}^{\rm 3D}$ experience significantly stronger dynamical heating, shortening their survival times below Hubble time. This effect is further amplified in more massive galaxies, where stronger tidal fields reduce the Jacobi radius and increase the cluster’s tidal filling. Adopting fallback-dependent natal kicks instead of the $\eta=1$ idealization would increase the metallicity contrast in initial BH fractions. The most massive BHs, formed via the direct stellar collapse, are subject to minimal kicks. Since high-mass BHs are more populated in metal-poor clusters, their retention fraction would be higher, leading to a further amplification of the mass-loss rate ratio $\dot{M}_{\rm MP}/\dot{M}_{\rm MR}$ shown in \figref{fig:Mdot}. Observational hints of top-heavy initial mass functions \citep{Marks2012} in low-metallicity environments would further exacerbate $\dot{M}_{\rm MP}/\dot{M}_{\rm MR}$ by seeding even more massive BHSubs, making dissolution likely even in initially tidally under-filling systems \citep{Chatterjee2017,Giersz2019}.

\subsection{Caveats}\label{sec:Caveats}

\citet{Gieles2023} simulated clusters that correspond to the low-$N$, low-density regime of GC evolution, where the impact of BH-driven dynamical heating on mass loss is enhanced. However, more massive and denser GCs often remain in an expansion-dominated phase over Hubble time, where the relaxation timescale, rather than BH heating, primarily regulates the mass-loss rate. Consequently, extrapolation of Equation~\ref{eq:MassLoss} to higher-$N$, higher-density clusters should be approached with caution, as the resulting metallicity dependence of the mass-loss rate may be weaker than inferred from the lower-density models.

Although our simulations roughly reproduce the initial densities of MW GCs, our models have lower $N$ than typical MW GCs. This limitation arises from current computational constraints: fully resolved \Nbody simulations with realistic stellar numbers, binary fractions, and remnant populations remain extremely demanding. The extent of the depleted region in $R_{\rm g}^{\rm 3D}$-[Fe/H] space is sensitive to the initial $N$: clusters with more typical $N$ would dissolve more slowly, since GC dissolution times scale approximately with the half-mass relaxation time. Consequently, the resulting depletion region would be smaller, implying that the relative contribution of BH dynamics to the depletion of inner-galaxy metal-poor GCs may be reduced compared to large-scale galactic processes such as hierarchical assembly. Future studies employing higher-$N$ models and dense initial configurations will be essential to quantitatively assess the strength of metallicity-dependent BH dynamics in shaping GC survival across diverse galactic environments.

\begin{figure*}
  \centering\includegraphics[width=1\linewidth]{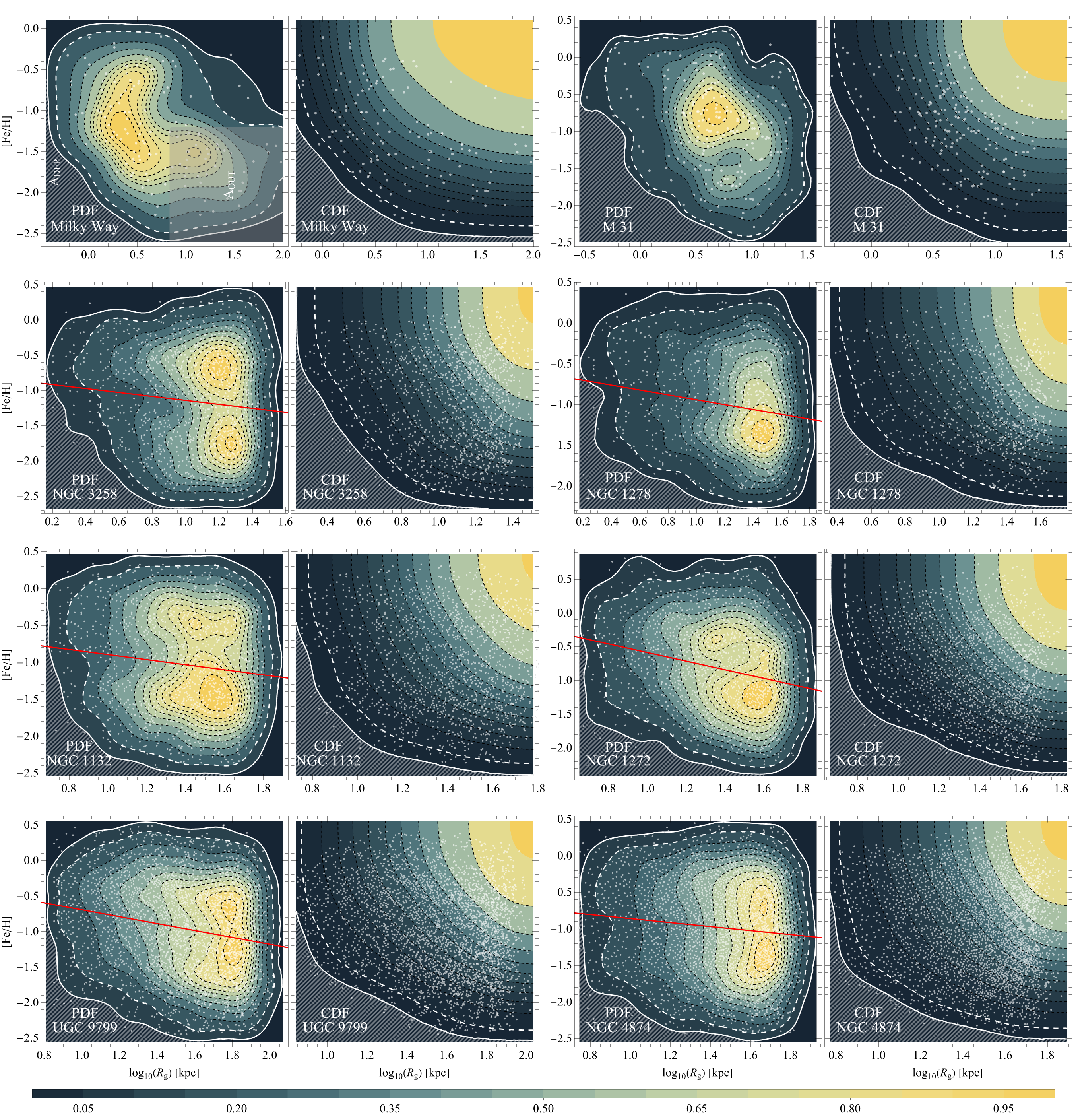}
  \caption{CDFs and scaled PDFs for eight galaxies in the $\Rg$-[Fe/H] parameter space. Dashed isodensity contours represent boundaries enclosing 5-95$\%$ of all data in $10\%$ increments. White contours (dashed for $95\%$ and solid for $99\%$, respectively) mark outer boundaries. Gray dots denote the distribution of GCs in each host galaxy. The hatched area highlights the depleted region of metal-poor GCs (\(\Adep\)), while the shaded area in the MW panel indicates the outer non-depleted region (\(\Aout\); see \secref{sec:Statistic}). Red solid lines show the least-squares fitted metallicity gradient (\citealt{Harris2023}; see \secref{sec:Assembly}). The CDFs are derived from the clustering results of the DBSCAN algorithm, with low-density regions excluded and classified as noise.}
  \label{fig:8galaxies}
\end{figure*}

\begin{figure*}
  \centering\includegraphics[width=1\linewidth]{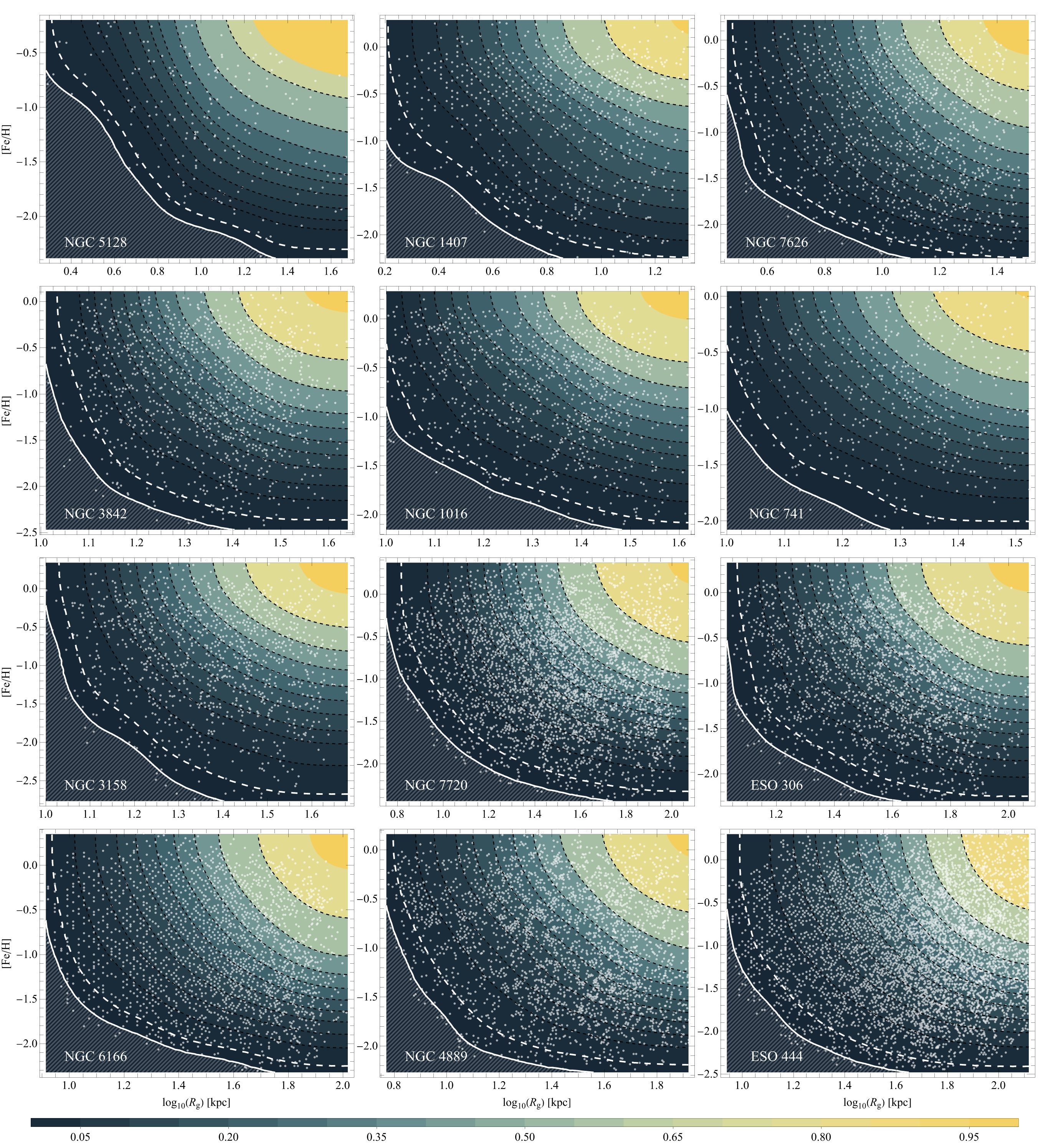}
  \caption{CDFs for nine galaxies in the $\Rg$-[Fe/H] parameter space, plotted from clustering results obtained via the DBSCAN algorithm. Contours are plotted similarly to those in \figref{fig:8galaxies}. The white contours mark boundaries that encompass 95 and 99$\%$ of the data, highlighting regions with very low probabilities of finding GCs, corresponding to approximate CDF values of 0.01 and 0.002. The hatched area highlights the depleted region of metal-poor GCs.}
  \label{fig:CDF}
\end{figure*}


\section{Results}\label{sec:results}

To investigate the spatial distribution of GC systems, we analyzed a sample of 37 host galaxies, encompassing both early-type and late-type galaxies. The results of this analysis are discussed in the following sections.

\subsection{The $\Rg$-[Fe/H] distribution of GC systems in host galaxies}

We constructed a two-dimensional parameter space using projected galactocentric distances ($\Rg$) and metallicity ([Fe/H]), where each GC occupies a point within this space. The list of galaxies, along with the associated references for their GC system data ($\Rg$ and [Fe/H]), is compiled in \tabref{tab:galaxies}. We utilize the multivariate (two-dimensional) probability density function (PDF) and cumulative distribution function (CDF) of GCs in the aforementioned parameter space as criteria to determine the spatial distribution patterns of metal-poor and metal-rich GCs in their host galaxies.

We employ Gaussian kernel density estimation (KDE) through the \texttt{scikit-learn} library in Python, specifically utilizing the \texttt{KernelDensity} estimator to derive both the PDF and CDF, with an automatic bandwidth selection method to optimize smoothing. The Gaussian kernel provides a smooth estimate of the underlying distribution, effectively capturing its structure. To visualize the distribution, we calculate isodensity contours based on the derived PDF and CDF. These contours are drawn at specific quantiles, representing boundaries that enclose certain percentages of the data. By analyzing these contours, we can identify regions where GCs are least likely to be found.

\figref{fig:8galaxies} illustrates the CDF and scaled PDF for eight galaxies, along with their corresponding isodensity contours. The PDFs for NGC 3258, NGC 1132, NGC 1272, UGC 9799, and NGC 4874 exhibit two distinct peaks: one associated with low metallicity and the other with high metallicity, reflecting the known bimodal metallicity distribution of GCs. As we move away from the peaks, the PDF asymptotically approaches zero, indicating regions where the probability of finding GCs is extremely low. \figref{fig:8galaxies} clearly demonstrates that the probability of locating the metal-poor population in the inner parts of the host galaxies is very low, with a specific trend: as the metallicity of GCs decreases, there is a corresponding reduction in the probability of their presence at smaller $\Rg$. This pattern was consistently observed in the PDFs of all 37 galaxies examined in this study.

This result mirrors the triangular depleted region observed in the mean Galactocentric distances, $a$, versus [Fe/H] diagram for the MW GC population, as discussed in \citetalias{Rostami2024-metallicity}. It is noteworthy that using $R_{\rm g}^{\rm 3D}$ or mean galactocentric distances, rather than $\Rg$, would amplify the observed depletion pattern of metal-poor GCs. However, the lack of three-dimensional data for most galaxies constrained this analysis. Projection effects may account for the apparent presence of metal-poor GCs in the inner regions of galaxies. Utilizing three-dimensional distances would push these clusters outward in parameter space, making the depleted region more distinct.

The CDF provides further insights into the spatial distribution of GCs. To construct the CDF, we applied the Density-Based Spatial Clustering of Applications with Noise (DBSCAN) algorithm through the \texttt{scikit-learn} library in Python, an effective unsupervised machine-learning technique that identifies clusters based on density variations (see, e.g., \citealt{Castro-Ginard2018}). DBSCAN excels at distinguishing dense core regions from sparse, low-probability points labeled as noise; in our implementation, the neighborhood radius ($\epsilon$) was defined as the average width of a grid cell in a uniform 10×10 partitioning of the parameter space. We set the minimum number of points required to form a cluster to $\sim1\%$ of the total dataset, excluding low-density regions below this threshold as noise. By systematically filtering out these areas, DBSCAN makes the depleted region of host galaxies more pronounced, while also mitigating the projection effects that could obscure the intrinsic spatial structure of GCs. The CDFs presented in \figref{fig:8galaxies} further highlight the marked scarcity of metal-poor GCs in the inner regions of these galaxies, following a distinct triangular pattern.

\figref{fig:CDF} shows the CDF for nine additional galaxies, highlighting the sparse distribution of metal-poor GCs in the inner regions across all galaxies. The solid white contours, which encompass $99\%$ of the data (corresponding to approximate CDF values of 0.002), outline large inner areas with a low probability of finding metal-poor GCs.


\subsection{Statistical tests}\label{sec:Statistic}

To systematically assess the sparsity of metal-poor GCs in the inner regions of host galaxies, we now apply a series of statistical tests. These methods allow us to quantify the significance of the depleted region and determine whether the observed patterns are statistically robust rather than a random fluctuation.


\subsubsection{Total probability ratio}\label{sec:TotP}

As a first step in our statistical analysis, we introduce a probability ratio metric that compares the observed probability of metal-poor GCs within the depleted region to their expected probability under the assumption that they follow the spatial distribution of metal-rich GCs. The depleted region (\(\Adep\)) is defined as the area outside the contour enclosing $99\%$ of the data in the PDF, situated in the inner regions of the host galaxy (indicated by the hatched area in the PDFs of \figref{fig:8galaxies}). The total probability of finding a metal-poor GC in the depleted region, as derived from the observed PDF, is denoted as \(\Pdep\). To estimate the expected probability, we assume that metal-poor GCs follow the same spatial distribution as metal-rich ones. To model this scenario, we construct a flipped probability density function (\(\PDFf\)), which mirrors the distribution of metal-rich GCs. We then compute the total probability based on \(\PDFf\) in \(\Adep\), denoted as \(\Pfdep\). Finally, we define the total probability ratio as
\begin{equation}
\PRdep = \frac{\Pfdep}{\Pdep}.
\end{equation}
For comparison, we repeat the same procedure for the non-depleted region (\(\Aout\)), defined as the rectangular area beyond the outer edge of the \(\Adep\), extending from the maximum depletion radius and restricted to GCs with [Fe/H] \(< -1.2\) (shown as the shaded area in the MW panel of \figref{fig:8galaxies}). The corresponding probability ratio is denoted as \(\PRout\).

The results, presented in \tabref{tab:galaxies}, show that \(\PRdep\) is consistently much larger than 1 across all galaxies, with an average value of 7.69. This suggests that if metal-poor GCs in the inner regions followed the same distribution as metal-rich ones, their probability density would be approximately eight times higher than what is observed. In contrast, the values of \(\PRout\) are close to 1 for most galaxies, with an average of 1.36, indicating that outside the depleted region, the probability density of metal-poor and metal-rich GCs is nearly identical. These results provide statistical evidence that the inner regions of host galaxies are indeed areas of anomalously low probability for metal-poor GCs. The probability density of metal-poor GCs within these regions differs significantly from that of their metal-rich counterparts, reinforcing the conclusion that metal-poor GCs are preferentially rare in the inner galactic regions.

\subsubsection{Nearest neighbor distance ratio}

While the total probability ratio provides a global measure of sparsity, a more localized approach can offer further insights. To this end, we examine the nearest neighbor distances of GCs within the depleted region, providing a direct measure of clustering. This allows us to compare the local probability density of metal-poor GCs with the expected distribution if they followed the spatial pattern of metal-rich GCs. We first compute the mean nearest neighbor distance for metal-poor GCs inside the depleted region. To establish a reference distribution, we assume that metal-poor GCs follow the same spatial pattern as metal-rich ones and compute the corresponding mean nearest neighbor distance using the flipped probability density function, \( \PDFf \). The ratio of these reference and observed values is denoted as \( \NNDRdep \). A value significantly smaller than 1 indicates that metal-poor GCs in the depleted region are much more sparsely distributed than metal-rich GCs. We repeat the same calculation for \( \Aout \), defining the corresponding ratio as \( \NNDRout \).

The results, summarized in \tabref{tab:galaxies}, reveal a significant difference in clustering behavior between the depleted and non-depleted regions. In the depleted region, \( \NNDRdep \) is consistently much smaller than 1 across all galaxies, with an average value of 0.33; for several galaxies, this value is zero, indicating that the depleted region is either completely devoid of metal-poor clusters or contains only one cluster, which leads to an infinite nearest neighbor distance. This indicates that the typical nearest neighbor distance for metal-poor GCs in the depleted region is approximately three times larger than expected under a metal-rich-similar distribution, reinforcing the conclusion that metal-poor GCs are much more sparsely distributed in the central regions of galaxies. In contrast, \( \NNDRout \) is close to 1 for most galaxies, with an average of 0.93, suggesting that outside the depleted region, the clustering properties of metal-poor and metal-rich GCs are nearly identical.

\subsubsection{Chi-square test}

To further test whether the observed spatial depletion of metal-poor GCs is statistically significant, we apply a chi-square test. This allows us to compare the observed GC distribution with expectations derived from the assumed spatial pattern of metal-rich GCs. The phase space was divided into a \( 10 \times 10 \) grid, partitioning both \(\Rg\) and [Fe/H] into ten equal bins. Within \(\Adep\), we compared the observed and expected distributions of GCs, where the expected values were derived from \(\PDFf\). The goodness of fit between the observed and expected distributions was assessed using a chi-square (\(\chi^2\)) test, with the corresponding \( p \)-value (\(\pchi\)), which quantifies the statistical significance of the deviation. The null hypothesis states that the observed distribution of metal-poor GCs in the depleted region follows the same spatial pattern as metal-rich GCs. A small \(\pchi\) indicates that the null hypothesis should be rejected, indicating that the observed distribution significantly deviates from the expected one.

The results detailed in \tabref{tab:galaxies} show that for most galaxies, \(\pchi\) falls below the conventional threshold of 0.05, indicating a significant deviation between the observed and expected distributions. These results provide further statistical confirmation that metal-poor GCs in the inner regions exhibit a markedly different spatial distribution compared to metal-rich GCs. The average \(\pchi\) across all studied galaxies is 0.07, reinforcing the conclusion that the scarcity of metal-poor GCs in the inner regions is unlikely to be a random fluctuation. Instead, a physical effect is responsible for this depletion.

\subsubsection{Poisson distribution analysis}

In addition to the chi-square test, we use a Poisson distribution analysis to evaluate whether the observed number of metal-poor GCs in the depleted region is consistent with expectations. This provides an independent statistical check on the robustness of our findings. Using the total probability values obtained in \secref{sec:TotP}, we compute the expected and observed cluster counts by multiplying the total probability by the total number of GCs in each galaxy. The Poisson probability, \(\ppoisson\), quantifies the likelihood of obtaining the observed count given the expected value.

The results in \tabref{tab:galaxies} indicate that \(\ppoisson\) approaches zero for nearly all galaxies. This suggests that the observed number of metal-poor GCs in the depleted region is highly improbable if they follow the spatial pattern of metal-rich GCs. The low probability values reinforce the conclusion that the scarcity of metal-poor GCs in the inner regions is not a statistical fluctuation but a physical effect. It is important to note that the validity of the Poisson approach depends on sample size, as the distribution becomes less meaningful for large expected counts. However, since the number of metal-poor GCs in the depleted region is relatively small across all galaxies, the Poisson model remains appropriate.

\subsubsection{Kolmogorov-smirnov test}

Finally, we perform a Kolmogorov-Smirnov (K-S) test to compare the metallicity distributions of metal-poor and metal-rich GCs. This test helps to determine whether the two populations originate from the same parent distribution, providing further information on the significance of the depleted region. Since the K-S test is inherently one-dimensional, we used only the metallicity values of GCs for this analysis. The inner surface was defined by first determining the largest radius within which the depleted region pattern exists. We then considered all clusters with \(\text{[Fe/H]} < -1.2\) within this radius. The observed and expected metallicity distributions were then obtained as before, with the expected distribution derived from \(\PDFf\). We computed the K-S test statistic and corresponding \( p \)-value for this region, \(\pksdep\). The same procedure was applied to the outer region, corresponding to \(\Aout\), yielding \(\pksout\).

The results reveal a striking contrast between the inner and outer regions (\tabref{tab:galaxies}). For many galaxies, \(\pksdep\) is below the conventional 0.05 threshold, with an average value of 0.03, indicating that the distributions of metal-poor and metal-rich GCs in the inner regions are statistically distinct. This strongly suggests that the presence of the depleted region plays a key role in differentiating the two populations. In contrast, \(\pksout\) is greater than 0.05 for most galaxies, with an average value of 0.36. This implies that outside the depleted region, the distributions of metal-poor and metal-rich GCs are much more similar, and there is no strong statistical evidence to reject the hypothesis that they originate from the same parent distribution.

These findings reinforce the conclusion that the depletion of metal-poor GCs in the inner regions is not a random fluctuation but a real effect that significantly alters the distribution. The stark contrast between \(\pksdep\) and \(\pksout\) further confirms that the primary factor distinguishing metal-poor and metal-rich GCs in the inner galaxy is the presence of the depleted region, whereas in the outer galaxy, their distributions are more comparable.

\begin{figure}
  \centering
  \includegraphics[width=1\linewidth]{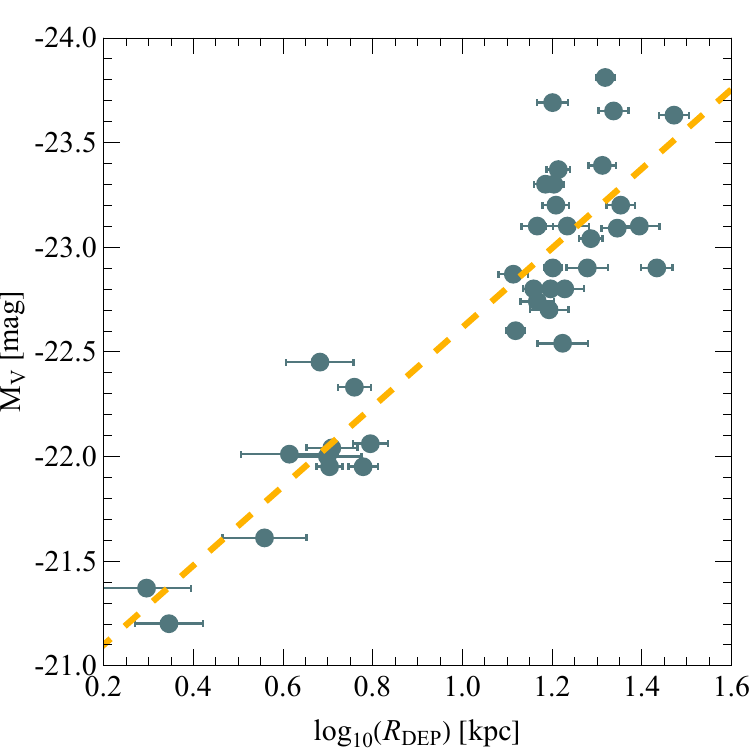}
  \caption{Mean $R_\mathrm{DEP}$ versus $V$-band absolute magnitude of host galaxies ($M_V$). Error bars indicate the standard deviation of $R_\mathrm{DEP}$. The fitted dashed line reveals the trend that more luminous galaxies tend to have larger $R_\mathrm{DEP}$ values, with a Pearson correlation coefficient of $0.9$.}
  \label{fig:Rvoid}
\end{figure}

\subsection{The influence of host galaxy luminosity on depletion morphology}

To quantify the influence of the host galaxy's tidal field on the morphology of the depleted region, we define the depopulation radius, $R_\mathrm{DEP}$, in the $\Rg$-[Fe/H] diagram. This parameter is defined as the distance at which the probability of encountering a cluster with [Fe/H] = -2.0 surpasses $1\%$ in the host galaxies. In other words, $R_\mathrm{DEP}$ is the distance at which the CDF for GCs with [Fe/H] = -2.0 reaches 0.01. Mathematically, this can be expressed as
\begin{equation}\label{eq:Rdep}
      f_\mathrm{CDF}(R_\mathrm{DEP}, [\mathrm{Fe/H}]=-2.0) = 0.01,
\end{equation}
where $f_\mathrm{CDF}$ represents the CDF of the GC distribution in the host galaxy. $R_\mathrm{DEP}$ serves as a proxy for the boundary beyond which metal-poor GCs ([Fe/H] $\geq$ -2.0) can persist against tidal disruption. We assume that this metric represents the edge of the depleted region. The integrated $V$-band absolute magnitude ($M_V$; see \tabref{tab:galaxies}) versus $R_\mathrm{DEP}$ for the complete set of 37 galaxies in our study is visualized in \figref{fig:Rvoid}. We used bootstrap resampling to determine $R_\mathrm{DEP}$ for each galaxy, generating 1000 iterations of galaxy's CDF while preserving the original GC count. $R_\mathrm{DEP}$ was calculated for each iteration, and the average values, with standard deviation as error bars, are shown in \figref{fig:Rvoid}. This method accounts for statistical uncertainties in $R_\mathrm{DEP}$. \figref{fig:Rvoid} shows a strong positive correlation between $R_\mathrm{DEP}$ and $M_V$, with a Pearson correlation coefficient of 0.9, suggesting that metal-poor GCs survive farther from more luminous galaxies, thereby shifting the survival boundary to larger $\Rg$. Although $R_\mathrm{DEP}$ is defined using [Fe/H] = -2.0 and $f_\mathrm{CDF}=0.01$, similar trends appear for other metallicities and thresholds.

Our initial hypothesis proposed a link between the morphology of the depleted region of metal-poor GCs and the total mass of their host galaxies. However, our results reveal a direct correlation with luminous mass. This result aligns with interpretations based on modified gravity theories or the $\Lambda$CDM framework. Interestingly, the correlation between $R_\mathrm{DEP}$ and $M_V$ indicates that the morphology of the metal-poor GC depleted region has potential as a novel distance indicator, offering an alternative method for measuring galactic distances. \figref{fig:Rvoid} showcases a heterogeneous sample of galaxies, encompassing Brightest Cluster Galaxies (BCGs) and ordinary field galaxies. Despite the distinct formation histories of these galaxy types, we did not separate them in this analysis. To develop a reliable method for galactic distance measurement, future studies may need to categorize galaxies based on their types and analyze them separately (\textcolor{blue}{Rostami-Shirazi et al.}, in prep.).

\subsection{Robustness of $R_\mathrm{DEP}$ calculations among limited central GC data in galaxies}

\begin{figure}
  \centering
  \includegraphics[width=1\linewidth]{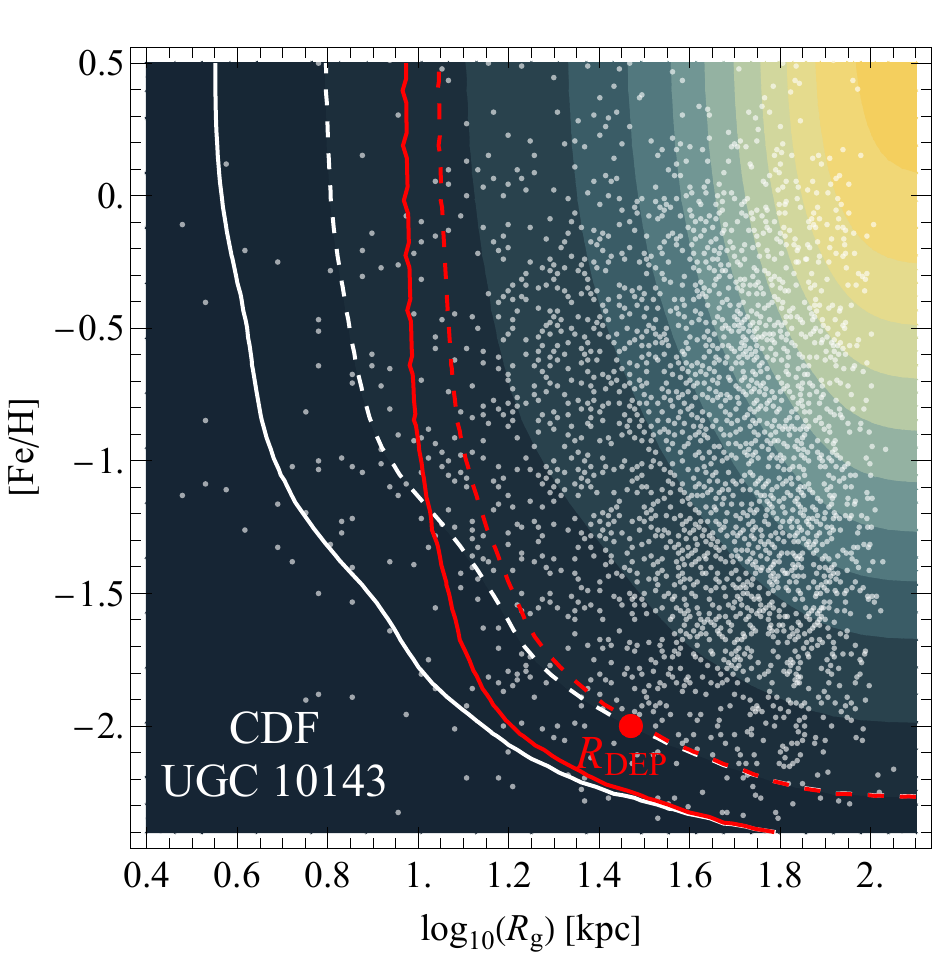}
  \caption{CDF for the galaxy UGC 10143 in the $\Rg$-[Fe/H] parameter space. White contours show the CDF for the full dataset (dashed line: $f_\mathrm{CDF} = 0.01$; solid line: $f_\mathrm{CDF} = 0.002$). Red contours represent the CDF calculated using only data with $\Rg > 10\kpc$ (same line style convention). The red dot marks the location of $R_\mathrm{DEP}$.}
  \label{fig:UGC10143}
\end{figure}

Studying GC systems presents a notable challenge, as data for many galaxies is typically available only from a few kpc outward. This limitation arises from the difficulty of detecting GCs against the bright background of the galaxy center, where the light often obscures their measurement. Consequently, detailed central GC distributions are generally restricted to the MW, M31, and a few Local Group galaxies. While this constraint could affect $R_\mathrm{DEP}$ calculations, our methodology demonstrates resilience. For instance, in the case of UGC 10143, where GC data is available for $\Rg>3\kpc$, we compared the CDF contours at 0.01 and 0.002 levels using the full dataset and a subset excluding data within 10 kpc. As shown in \figref{fig:UGC10143}, the contours converged at low metallicities, with $R_\mathrm{DEP}$ calculations differing by less than $1\%$ between the two datasets. Repeating this test for other galaxies consistently showed errors below a few percent, confirming that our method effectively accounts for the lack of central GC distribution data. Indeed, the pattern of depleted regions extends several $\kpc$ from the galaxy center. Even in the absence of GC distribution data in the central regions, the calculation of $R_\mathrm{DEP}$ remains robust.

The mass and characteristic radius of stellar systems (including galaxies) typically correlate with each other. It is therefore natural to expect that this mass-radius correlation would also apply to the total mass and radial extent of any early central star formation episode. Consequently, in more massive galaxies, the region where metal-rich central clusters are formed is expected to be more extended. While this provides a structural expectation, additional dynamical effects influence the survival and spatial distribution of clusters in these inner regions, depending on the mass of the host galaxy. However, our analysis may not fully capture the complex dynamics of inner galactic regions, where additional external effects such as rapidly varying tidal forces \citep{Gnedin1997}, tidal shocks from disk crossings \citep{Gnedin1999}, interactions with giant molecular clouds \citep{Gieles2006}, and spiral arm influences \citep{Gieles2007} can significantly contribute to GC dissolution. This caveat underscores the need for further investigation into these factors when interpreting our results for the inner galactic regions.

\subsection{Depleted regions and galactic assembly histories}\label{sec:Assembly}

\begin{figure*}
  \centering\includegraphics[width=1\linewidth]{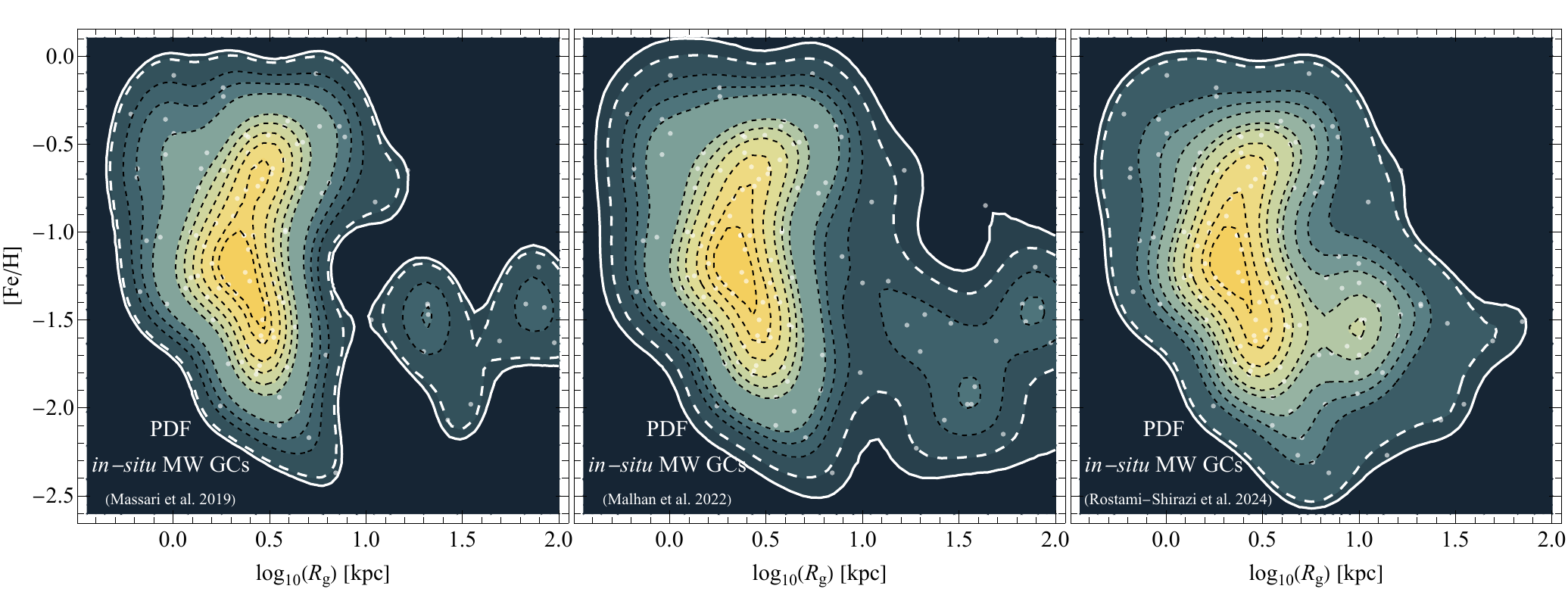}
  \caption{Scaled PDFs of MW GCs restricted to the insitu population, analogous to \figref{fig:8galaxies}. Exsitu GCs are excluded based on \citeauthor{massari2019} (\citeyear{massari2019}; left panel), \citeauthor{Malhan2022} (\citeyear{Malhan2022}; middle panel), and 
\citeauthor{Shirazi2023} (\citeyear{Shirazi2023}; right panel).}
  \label{fig:MW-insitu}
\end{figure*}

The assembly scenario for GC system formation posits that metal-poor GCs are primarily captured from satellite galaxies, resulting in a broader spatial distribution than that of metal-rich GCs. In addition, major mergers involving gas-rich progenitors can trigger the formation of new, centrally concentrated metal-rich clusters, while simultaneously redistributing preexisting metal-poor GCs to larger radii through angular momentum transfer \citep{Bekki2002,Li2004}. These mechanisms successfully account for the observed metallicity gradients with $\Rg$, where the average metallicity of GCs decreases as $\Rg$ increases. Red solid lines in \figref{fig:8galaxies} illustrate this trend, showing least-squares fits in the form
\begin{equation}\label{eq:redline}
      \mathrm{[Fe/H]} = a + b ~ \mathrm{log_{10}}(\Rg).
\end{equation}
\citet{Harris2023} demonstrated shallow gradients in mean metallicity for most galaxies in their sample, with a mean power-law exponent of $\langle b\rangle=-0.3$ across the entire dataset and, notably, no clear trend with galaxy luminosity. While the shallow metallicity gradient is well justified by both assembly and major merger scenarios, it appears insufficient to fully explain the observed triangular depleted region of metal-poor GCs in the inner parts of galaxies. Crucially, we observed a clear correlation between the morphology of this depleted region and the luminous mass of galaxies, a trend not present in the metallicity gradients.

To assess whether the observed depleted region is solely a signature of the assembly history, we constructed PDFs of the MW GC system, separating insitu formed clusters from exsitu accreted ones. A common approach for this classification is to distinguish GCs in the space of Galactic orbital energy and out-of-plane angular momentum, often associating them with known interacting events (e.g., \citealt{massari2019,Naidu2020,Malhan2022,Bonaca2025}). \figref{fig:MW-insitu} shows the PDF of the MW GCs restricted to the insitu population, based on three independent classification schemes. \citet{massari2019} associate $\sim 35\%$ of MW GCs with major mergers, including Gaia-Enceladus, the Sagittarius dwarf galaxy, the Helmi streams, and the Sequoia galaxy. \citet{Malhan2022} employed a clustering algorithm in action space to identify six distinct merger events, including Pontus. \citet{Shirazi2023} performed kinematic simulations to trace potential GC capture from 41 dwarf satellite galaxies, identifying 29 likely exsitu candidates. The PDFs derived from all three studies consistently reveal that, even when exsitu clusters are excluded, the triangular depleted region of metal-poor GCs remains clearly evident. Quantitatively, the value of $R_\mathrm{DEP}$ changes by less than $6\%$ compared to the analysis of the full GC sample. This suggests that the depleted region is not merely a byproduct of the Galaxy’s hierarchical assembly.

Although hierarchical assembly naturally predicts a broader spatial distribution for metal-poor GCs, it falls short of reproducing the distinct triangular morphology of the depleted region and its systematic dependence on galaxy luminosity. These features instead point toward an additional, internally driven mechanism operating within the GCs themselves. BH dynamics provides a plausible contributing factor to both the universality of the depletion pattern and its scaling with host galaxy mass. This mechanism likely acts in concert with large-scale processes, such as hierarchical accretion histories and environmental effects, in shaping the spatial metallicity distribution of GCs. A quantitative validation of this interpretation remains essential. Specifically, determining the slope of the $R_\mathrm{DEP}$–$M_V$ trend from simulations and comparing it with the observed trend (\figref{fig:Rvoid}) constitutes a key test, which will be the focus of a future study.

\section{Summary and conclusion}\label{sec:conclusion}

We analyzed the spatial distribution of GC systems in 37 host galaxies by examining their PDFs and CDFs in the $\Rg$-[Fe/H] parameter space. Our investigation revealed a universal triangular depleted region, marked by a distinct lack of metal-poor GCs in the inner parts of galaxies. This pattern demonstrates that GCs with lower metallicities appear further away from the galactic center, providing evidence that the dissolution time of GCs within galaxies is metallicity-dependent. This dependency is contingent upon a subsystem of BHs, BHSub, initially retained within the GC. The BHSub acts as an energy source, fueling the GC's stellar population through kinetic energy generated by few-body BH encounters. However, additional factors such as the hierarchical build-up of galaxies and tidal interactions must also be considered to fully explain the observed radial metallicity trends.

The more massive and less expanded BHSub in metal-poor GCs fosters more intense few-body encounters, injecting more kinetic energy into the stellar population. This extra energy, coupled with the strong tidal field in the galactic central regions, accelerates the dissolution of lower-metallicity GCs on timescales shorter than the host galaxy's age. As the $\Rg$ increases, clusters experience a weaker galactic tidal field, reducing the impact of this excess internal energy on metal-poor GCs' evaporation rates, thus enabling their survival to the present day. In more massive galaxies, stronger tidal forces push surviving metal-poor GCs farther from the center, broadening the triangular depleted region in $\Rg$-[Fe/H] space. Our analysis reveals a strong positive correlation between the morphology of the metal-poor GCs depleted region and the luminous mass (absolute magnitude; $M_V$) of the host galaxies, as quantified by a Pearson correlation coefficient of $0.9$. More massive galaxies consistently exhibit more extended depleted regions. This correlation can serve as a novel method to estimate the galaxy distances.

These findings underscore the essential role of initial BH retention in the long-term evolution of GCs, suggesting limitations for scenarios in which BHs receive substantial natal kicks. In such models, the absence of a BHSub would eliminate the metallicity-dependence of cluster dissolution, and consequently fail to reproduce the triangular depleted region observed in the $\Rg$–[Fe/H] diagram.


\bibliographystyle{aa}

\clearpage        
\onecolumn        
\appendix
\section{Galaxy sample and data sources}\label{sec:Appendix}

Properties of the studied galaxies, data sources for their GC systems, and statistical analyses evaluating the depletion of metal-poor GCs are summarized in \tabref{tab:galaxies}.

\FloatBarrier     
\begin{table}[!htbp]   
\centering
\caption{Properties of the studied galaxies and their GC systems.}
\begin{tabular}{ccccccccccc}
    \hline 
  Galaxy & $M_V$ & $\overset{\large \sim}{\small{\mathrm{P}}}_\mathrm{DEP}$ & $\overset{\large \sim}{\small{\mathrm{P}}}_\mathrm{OUT}$ & $\overset{\large \sim}{\small{\mathrm{NND}}}_\mathrm{DEP}$ & $\overset{\large \sim}{\small{\mathrm{NND}}}_\mathrm{OUT}$ & $p_\mathrm{KS-DEP}$ & $p_\mathrm{KS-OUT}$ & $p_\mathrm{\upchi^2}$ & $p_\mathrm{Poisson}$ & $R_\mathrm{DEP}$ \\
     & (mag) &  &  &  &  &  &  &  &  & ($\kpc$) \\
    \hline 

M 81\tablefootmark{2} & $-21.20$\tablefootmark{1} & 7.88 & 1.20 & 0 & 0.48 & 0.00 & 0.80 & 0.12 & 0.00 & $2.22\pm1.19$ \\

Milky Way\tablefootmark{a}\tablefootmark{4}\tablefootmark{5} & $-21.37$\tablefootmark{3} & 7.91 & 0.39 & 0 & 1.65 & 0.05 & 0.01 & 0.03 & 0.00 & $1.98\pm1.25$ \\

M 31\tablefootmark{8} & $-21.61$\tablefootmark{6}\tablefootmark{7} & 17.04	& 0.66 & 0 & 1.40 & 0.12 & 0.87 & 0.08 & 0.00 & $3.62\pm1.24$ \\

NGC 3258\tablefootmark{9} & $-21.95$\tablefootmark{9} & 5.12 & 1.09 & 0.46 & 0.98 & 0.01 & 0.89 & 0.00 & 0.00 & $5.06\pm1.07$ \\

NGC 3268\tablefootmark{9} & $-21.95$\tablefootmark{9} & 4.67 & 1.20 & 0.27 & 0.97 & 0.17 & 0.36 & 0.02 & 0.00 & $6.01\pm1.08$ \\

NGC 5322\tablefootmark{9} & $-22.00$\tablefootmark{9} & 3.80 & 0.95 & 0 & 1.11 & 0.22 & 0.02 & 0.22 & 0.08 & $5.00\pm1.19$  \\

NGC 5128\tablefootmark{11}\tablefootmark{12} & $-22.01$\tablefootmark{6}\tablefootmark{10} & 6.52 & 1.23 & 0.55 & 0.97 & 0.00 & 0.07 & 0.22 & 0.00 & $4.12\pm1.28$ \\

NGC 1407\tablefootmark{9} & $-22.04$\tablefootmark{9} & 5.17 & 2.02 & 0.30 & 0.71 & 0.04 & 0.06 & 0.03 & 0.00 & $5.12\pm1.14$  \\

NGC 7626\tablefootmark{9} & $-22.06$\tablefootmark{9} & 2.71 & 1.03 & 0.67 & 0.98 & 0.04 & 0.30 & 0.36 & 0.00 & $6.24\pm1.09$ \\

 NGC 5557\tablefootmark{9} & $-22.33$\tablefootmark{9} & 10.50 & 1.26 & 0 & 0.88 & 0.04 & 0.07 & 0.41 & 0.00 & $5.75\pm1.09$  \\

M 104\tablefootmark{13} & $-22.45$\tablefootmark{2} & 6.37 & 0.73 & 0 & 1.45 & 0.07 & 0.48 & 0.17 & 0.00 & $4.82\pm1.19$ \\

 NGC 1278\tablefootmark{9} & $-22.54$\tablefootmark{6}\tablefootmark{9} & 3.04 & 0.60 & 0.57 & 1.34 & 0.07 & 0.50 & 0.05 & 0.00 & $16.75\pm1.14$  \\

NGC 1600\tablefootmark{14} & $-22.60$\tablefootmark{6} & 6.15 & 2.42 & 0.39 & 0.67 & 0.00 & 0.00 & 0.00 & 0.00 & $13.15\pm1.05$ \\

NGC 4914\tablefootmark{14} & $-22.70$\tablefootmark{6} & 20.51 & 4.55 & 0 & 0.43 & 0.02 & 0.09 & 0.01 & 0.00 & $15.62\pm1.10$  \\

NGC 1275\tablefootmark{9} & $-22.74$\tablefootmark{9} & 6.88 & 0.98 & 0.28 & 0.95 & 0.00 & 0.70 & 0.01 & 0.00 & $14.71\pm1.09$ \\

NGC 533\tablefootmark{14} & $-22.80$\tablefootmark{6} & 10.35 & 2.28 & 0.25 & 0.48 & 0.00 & 0.11 & 0.00 & 0.00 & $14.43\pm1.06$  \\

NGC 777\tablefootmark{14} & $-22.80$\tablefootmark{6} & 30.63 & 2.48 & 0.35 & 0.65 & 0.00 & 0.00 & 0.00 & 0.00 & $15.75\pm1.06$ \\

NGC 57\tablefootmark{14} & $-22.80$\tablefootmark{6} & 6.99 & 1.36 & 0.60 & 0.86 & 0.00 & 0.05 & 0.00 & 0.00 & $16.93\pm1.10$  \\

NGC 1132\tablefootmark{9} & $-22.87$\tablefootmark{9} & 3.87 & 0.91 & 0.45 & 1.06 & 0.00 & 0.89 & 0.00 & 0.00 & $13.00\pm1.08$ \\

NGC 3842\tablefootmark{14} & $-22.90$\tablefootmark{6} & 7.20 & 1.51 & 0.22 & 0.80 & 0.00 & 0.00 & 0.00 & 0.00 & $15.93\pm1.05$  \\

NGC 1016\tablefootmark{14} & $-22.90$\tablefootmark{6} & 8.60 & 2.02 & 0.56 & 0.73 & 0.00 & 0.12 & 0.00 & 0.00 & $19.01\pm1.11$ \\

UGC 10143\tablefootmark{9} & $-22.90$\tablefootmark{9} & 3.13 & 0.83 & 0.67 & 1.10 & 0.04 & 0.26 & 0.09 & 0.00 & $
27.16\pm1.08$  \\

ESO 509-G008\tablefootmark{9} & $-23.04$\tablefootmark{9} & 9.46 & 0.83 & 0.17 & 1.26 & 0.00 & 0.79 & 0.00 & 0.00 & $19.35\pm1.06$ \\

NGC 1272\tablefootmark{9} & $-23.09$\tablefootmark{9} & 5.54 & 0.81 & 0.51 & 1.15 & 0.00 & 0.95 & 0.00 & 0.00 & $22.17\pm1.08$  \\

NGC 410\tablefootmark{14} & $-23.10$\tablefootmark{6} & 11.67 & 1.81 & 0 & 0.56 & 0.00 & 0.63 & 0.26 & 0.00 & $14.70\pm1.08$ \\

NGC 741\tablefootmark{14} & $-23.10$\tablefootmark{6} & 11.73 & 1.69 & 0 & 0.76 & 0.01 & 0.01 & 0.31 & 0.00 & $17.16\pm1.12$  \\

NGC 2340\tablefootmark{14} & $-23.10$\tablefootmark{6} & 5.95 & 1.61 & 0.40 & 0.53 & 0.00 & 0.89 & 0.19 & 0.00 & $24.83\pm1.11$ \\

NGC 4839\tablefootmark{14} & $-23.20$\tablefootmark{6} & 10.07 & 1.81 & 0.16 & 0.84 & 0.00 & 0.00 & 0.00 & 0.00 & $16.19\pm1.07$  \\

UGC 9799\tablefootmark{9} & $-23.20$\tablefootmark{9} & 5.44 & 0.98 & 0.44 & 1.02 & 0.00 & 0.81 & 0.00 & 0.00 & $22.57\pm1.08$ \\

NGC 4073\tablefootmark{14} & $-23.30$\tablefootmark{6} & 10.12 & 1.89 & 0.12 & 0.76 & 0.00 & 0.00 & 0.03 & 0.00 & $15.33\pm1.06$  \\

NGC 3158\tablefootmark{14} & $-23.30$\tablefootmark{6} & 4.64 & 1.84 & 0.23 & 0.79 & 0.00 & 0.00 & 0.01 & 0.00 & $16.01\pm1.05$ \\

NGC 4874\tablefootmark{9} & $-23.37$\tablefootmark{9} & 3.70 & 0.92 & 0.50 & 1.04 & 0.01 & 0.90 & 0.00 & 0.00 & $16.38\pm1.06$  \\

NGC 7720\tablefootmark{9} & $-23.39$\tablefootmark{9} & 3.70 & 0.83 & 0.61 & 0.99 & 0.00 & 0.75 & 0.00 & 0.00 & $20.54\pm1.07$ \\

ESO 306-G017\tablefootmark{9} & $-23.63$\tablefootmark{9} & 4.63 & 1.01 & 0.65 & 1.05 & 0.00 & 0.19 & 0.00 & 0.00 & $29.67\pm1.08$  \\

NGC 6166\tablefootmark{9} & $-23.65$\tablefootmark{9} & 3.61 & 0.84 & 0.63 & 1.07 & 0.00 & 0.56 & 0.01 & 0.00 & $21.75\pm1.08$ \\

NGC 4889\tablefootmark{9} & $-23.69$\tablefootmark{9} & 7.27 & 0.84 & 0.47 & 1.06 & 0.04 & 0.26 & 0.00 & 0.00 & $15.91\pm1.08$ \\

ESO 444-G046\tablefootmark{9} & $-23.81$\tablefootmark{9} & 2.14 & 1.07 & 0.65 & 1.02 & 0.07 & 0.01 & 0.00 & 0.00 & $20.84\pm1.05$ \\

    \hline
Mean &  & 7.69 & 1.36 & 0.33 & 0.93 & 0.03 & 0.36 & 0.07 & 0.00 &  \\
    \hline
\end{tabular}
\tablefoot{
\textit{Columns:} (1) galaxy names and the associated references for their GC system data, (2) integrated $V$-band absolute magnitudes ($M_V$), (3) total probability ratio in the depleted region ($\PRdep$), (4) total probability ratio in the outer region ($\PRout$), (5) mean nearest neighbor distance ratio in the depleted region ($\NNDRdep$), (6) mean nearest neighbor distance ratio in the outer region ($\NNDRout$), (7) p-value from the Kolmogorov-Smirnov (K-S) test for the inner regions ($p_{\mathrm{KS-DEP}}$), (8) p-value from the K-S test for the outer regions ($p_{\mathrm{KS-OUT}}$), (9) p-value from the chi-square test in the depleted region ($p_{\mathrm{\upchi^2}}$), (10) Poisson probability for the depleted region ($p_{\mathrm{Poisson}}$), and (11) mean depopulation radius ($R_{\mathrm{DEP}}$) with its standard deviation, obtained from 1000 CDF resampling. 

\textit{References:} \tablefoottext{1} {\citet{Bruns2012};} \tablefoottext{2} {\citet{Jun2013};} \tablefoottext{3} {\citet{Hawthorn2016};} \tablefoottext{4} {\citet{Kruijssen2019};} \tablefoottext{5} {\citet{Baumgardt2019mean};} \tablefoottext{6} {\citet{Vaucouleurs1991};} \tablefoottext{7} {\citet{Komiyama2018};} \tablefoottext{8} {\citet{Galleti2004};} \tablefoottext{9} {\citet{Harris2023};} \tablefoottext{10} {\citet{Crnojevi2016};} \tablefoottext{11} {\citet{Beasley2008};} \tablefoottext{12} {\citet{Woodley2007};} \tablefoottext{13} {\citet{Alves-Brito2011};} \tablefoottext{14} {\citet{Hartman2023}}. \tablefoottext{a} {For the Milky Way (MW) analysis, the semi-major axis measurements of GCs were employed instead of $\Rg$.}
}
\label{tab:galaxies}
\end{table}


\label{lastpage}
\end{document}